\documentstyle[prb,preprint,aps,psfig]{revtex}

\newcommand{\be}{\begin{equation}}
\newcommand{\ee}{\end{equation}}
\newcommand{\bt}{\begin{tabular}}
\newcommand{\et}{\end{tabular}}

\begin{document}
\draft
\title{Off-equilibrium dynamics in finite dimensional spin glass models}

\author{J. Kisker, L. Santen, M. Schreckenberg}
\address{Institut f\"ur Theoretische Physik, Universit\"at zu K\"oln,
         D-50937 K\"oln, Germany\\ and Institut f\"ur Theoretische
         Physik - FB 10, Universit\"at Duisburg, D-47048 Duisburg, Germany}

\author{H. Rieger}
\address{Institut f\"ur Theoretische Physik, Universit\"at zu K\"oln,
         D-50937 K\"oln, Germany\\ and HLRZ, Forschungszentrum J\"ulich,
         D-52425 J\"ulich, Germany}

%
%
\maketitle

\begin{abstract}
The low temperature dynamics of the two- and three-dimensional Ising
spin glass model with Gaussian couplings is investigated via extensive
Monte Carlo simulations. We find an algebraic decay of the remanent
magnetization. For the autocorrelation function
$C(t,t_w)=[\langle S_i(t+t_w)S_i(t_w)\rangle]_{av}$
a typical aging scenario with a $t/t_w$ scaling is established.
Investigating spatial correlations
we find an algebraic growth law $\xi(t_w)\sim t_w^{\alpha(T)}$ of the average
domain size. The spatial correlation function
$G(r,t_w)=[\langle S_i(t_w)S_{i+r}(t_w)\rangle^2]_{av}$
scales with $r/\xi(t_w)$.
The sensitivity of the correlations in the spin glass phase with respect
to temperature changes is examined by calculating a time dependent
overlap length. In the two dimensional model
we examine domain growth with a new method: First we determine the exact
ground states of the various samples (of system sizes up to $100\times 100$)
and then we calculate the correlations between this state and the states
generated during a Monte Carlo simulation.

\end{abstract}

\pacs{75.10Nr, 75.50Lh, 75.40Mg}
\section{Introduction}
In spin glasses \cite{binderyoung} below the transition temperature
characteristic non-equilibrium phenomena can be observed \cite{mcreview}.
The typical
experiment where these phenomena are encountered is the measurement
of the (thermo-)remanent magnetization. The procedure is the following:
A spin glass sample is rapidly
cooled within a magnetic field to a temperature below the transition
temperature $T_g$ and then the field is switched off after a certain waiting
time $t_w$. The striking observation then is that the decay of the
magnetization $M(t)$ is found to depend on the waiting time $t_w$ even on
laboratory time scales, a phenomenon called aging \cite{lundgren}. Aging is
not restricted to spin glasses and has also been found in other disordered
or amorphous systems such as polymers \cite{polymers}, high temperature
superconductors \cite{hightc} and charge-density wave systems \cite{CDW}
where certain quantities show a characteristic history dependence.

Several attempts have been made to explain this behavior theoretically.
However, it has not been possible up to now to determine the non-equilibrium
dynamics of short-range, finite-dimensional spin glass models starting
from a microscopic Hamiltonian analytically.
Thus one depends on phenomenological models
which are either of a hierarchical type where the relaxation process
is described by diffusion in a tree like structure of phase-space
\cite{sibani,bouchaud} or scaling theories
which consider the energetically low lying excitations in
real space which are supposed to be connected cluster of reversed spins
\cite{fh,koper}. An important ingredience of the latter theories
are domain growth laws which determine the growth of the average domain size
in dependence of the waiting time. Experimentally this quantity cannot be
measured whereas in Monte Carlo (MC) Simulations one has direct access to all
quantities of interest. Thus MC Simulations are an important touchstone for
phenomenological theories and their underlying assumptions.

In this paper we present the results of large scale MC simulations of the
two (2D)- and three-dimensional (3D) Ising spin glass model with nearest
neighbour interactions and a {\em continuous} bond distribution. The focus
is on the three dimensional
model whose low-temperature dynamics is examined in detail by
calculating correlation functions in time {\em and} space. In earlier
works \cite{andersson,rieger+-j} mainly the case of
a {\em binary} bond distribution was studied and the focus was on the
time-dependent {\it auto-} correlations.
Thus by also calculating correlations in space one might hope to
gain a deeper insight into the aging process and a more direct check of the
phenomenological theories.
In contrast to the two dimensional model the 3D-model is known to
have a finite spin glass transition temperature of $T_g\sim 0.9$
\cite{bhatttc,mcmillan,braytc} for Gaussian couplings.
The non-equilibrium dynamics of the two-dimensional model has been investigated
in a previous paper \cite{bast} and an interrupted aging scenario was
reported there which means that the aging process is interrupted as soon
as the waiting time exceeds the equilibration time of the system which is
finite for non-vanishing temperatures in 2D. This led to
the conclusion that a finite transition temperature is not necessary to
observe aging effects. Even in simpler models without frustration
\cite{kis} or disorder \cite{pspin,cug} aging was found.
In the present paper we present the results of a new method to identify
domains in the two dimensional model. In systems whose ground state is not
known, this is a non-trivial problem and in the previous work \cite{bast}
as well as in the three-dimensional model a replica method has been used.
However, it is possible to compute the ground state of the 2D model
for fairly large system sizes which is used to calculate the average, time
dependent domain size and we compare the results of both methods.

The outline of the paper is as follows: In Sec. \ref{sectionmodel}
the three-dimensional spin glass model is introduced. In Sec.
\ref{sectionaging} and \ref{sectionspatial3d} the simulation results
of the autocorrelation function and the spatial correlations
are given. In Sec. \ref{sectionspatial2d} the spatial correlations
in the two-dimensional spin glass model are examined.

\section{The three-dimensional spin glass model}
\label{sectionmodel}
We consider the three-dimensional Ising spin glass with
nearest-neighbour interactions whose Hamiltonian is given by

\begin{equation}
  H=-\sum_{\langle ij \rangle} J_{ij} S_i S_j \; .
\label{hami}
\end{equation}
The $S_i= \pm 1$ are Ising spins on a $L \times L \times L$ simple
cubic lattice and the random interaction strengths $J_{ij}$ are drawn
from a Gaussian distribution
\begin{equation}
  P(J_{ij})=\frac{1}{\sqrt{2\pi}}\exp\left( -\frac{J_{ij}^2}{2}\right)
  \; ,
\label{distribution}
\end{equation}
with zero mean and variance one. We used single-spin-flip Glauber dynamics
where each spin is flipped with probability
\begin{equation}
  w(-S_i \rightarrow S_i) = \frac{1}{1+\exp(\Delta E/T)} \;,
\label{flipprob}
\end{equation}
$\Delta E$ being the energy difference between the new state with
$S_i=+1$ and the old state with $S_i=-1$. Time is measured in Monte
Carlo sweeps (MCS) through the whole lattice and periodic boundary
conditions were implied.

The simulations were performed below the transition temperature
$T_g=0.9 \pm 0.1$ \cite{bhatttc} in the spin glass
phase. Depending on temperature $T$, lattice sizes from $L=14$ to
$L=32$ were used. For lower temperatures smaller lattice sizes were
sufficient since the correlation length grows less rapidly for smaller
temperatures (see Sec. \ref{sectionspatial3d}). As the correlation
length is much smaller than the system sizes finite-size effects can
be excluded which we checked also explicitly by varying the system
sizes.

An Intel Paragon XP/S 10 parallel computer with 136 i860XP nodes and a
Parsytech GCel1024 transputer cluster with 1024 T805-processors have been used
for the simulations.
A single correlation function took about 16 hours of CPU time on
128 nodes of the Paragon system. On the Transputer cluster
a single processor is roughly 10 times slower than on the Intel
machine and thus one run took about 160 hours on 128 nodes there.

\section{The autocorrelation function in the 3D model}
\label{sectionaging}
A simple way to observe aging effects in the present model is to
calculate the autocorrelation function
\begin{equation}
  C(t,t_w)=\frac{1}{N}\sum_i[\langle S_i(t+t_w)S_i(t_w)\rangle ]_{av}
\label{autocorr}
\end{equation}
which measures the overlap of the spin configurations at times $t+t_w$
and $t_w$. $[\ldots ]_{av}$ indicates an average over different
realizations of the bond disorder and $\langle \ldots \rangle$ is a
thermal average, i.e. an average over different initial conditions and
realizations of the thermal noise. Initially the spins take on random values
$S_i=\pm 1$ corresponding to a quench from $T=\infty$ to the temperature
$T<T_g$ at which the simulation is run.

\begin{figure}
\psfig{file=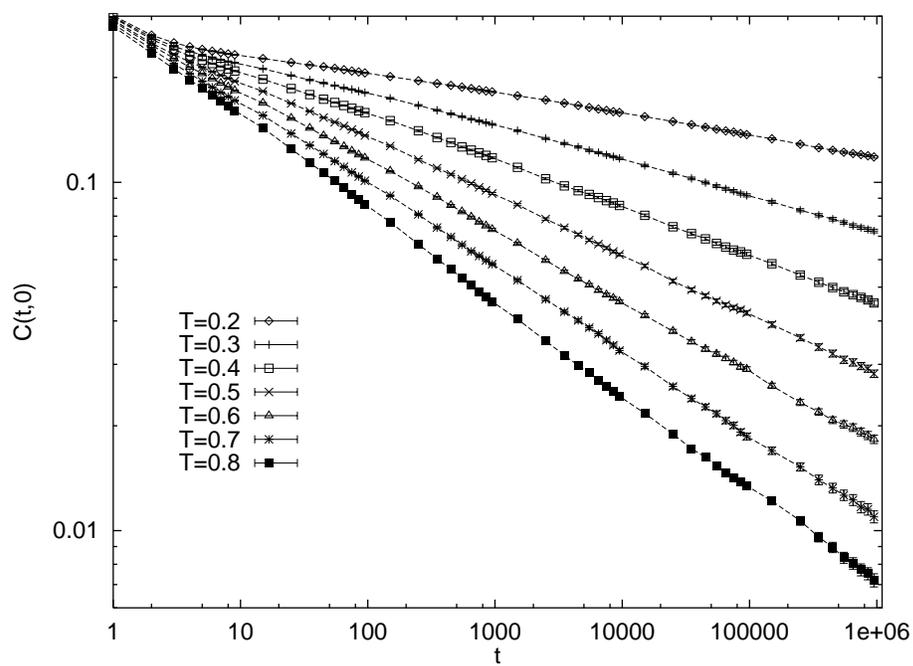,width=14cm}
\caption{The remanent magnetization of the 3D EA-model (\protect\ref{hami})
for temperatures \protect{$T=0.2$ to $T=0.8$} (from top to bottom).}
\label{fig1}
\end{figure}

\begin{figure}
\psfig{file=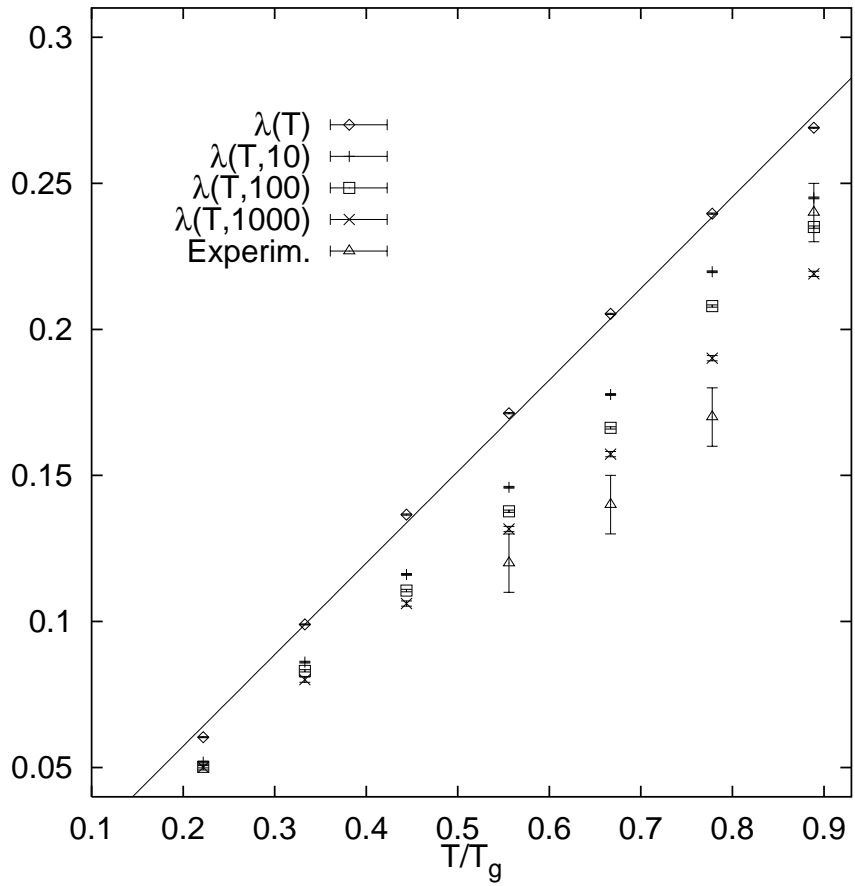,width=14cm}
\caption{The non-equilibrium exponent $\protect{\lambda(T,t_w)}$ of the 3D
EA-model. The straight line is a linear fit of $\protect{\lambda(T)}$ and
is a guideline for the eye only. The experimentally determined  exponents
are taken from [22].}
\label{fig2}
\end{figure}

First we discuss the quantity $C(t,0)$ which corresponds to the
remanent magnetization $M(t)$ after a quench from infinite magnetic field.
Note that for a symmetric bond distribution the fully magnetized state
($S_i =+1$ for all $i$) and a random initial configuration are completely
equivalent.
In Fig. \ref{fig1} the remanent magnetization is
shown for different temperatures in a log-log plot. One observes that after a
few time steps the decay of $C(t,0)$ clearly is algebraic. Fits to the
function
\be
  M(t) \sim t^{-\lambda(T)}
\label{remmag}
\ee
give the set of exponents $\lambda(T)$ which are shown in Fig.
\ref{fig2}. Temperatures were converted to ratios $T/T_g$
with $T_g=0.9$. We also tried logarithmic fits of the data as was proposed in
\cite{fh} within the droplet theory but they did not
give acceptable results. For better readability a linear fit
of $\lambda(T)$ is also shown. The exponents $\lambda(T)$ increase linearly
with temperature to a good approximation.

The remanent magnetization can also be measured experimentally. This can be
done by fully polarizing a sample in a magnetic field, switching off the
field and then measuring the decay of the time dependent magnetization.
In experiments with an amorphous, metallic spin
glass $\rm (Fe_{0.15}Ni_{0.85})_{75}P_{16}B_6Al_3$ \cite{granberg} an
algebraic decay of $M(t)$ was found for temperatures below $T_g$.
We read off the resulting exponents from figure 4(b) in \cite{granberg}
and included them in Fig. \ref{fig2}. Comparing these exponents with the
ones from the simulations one finds that there are quantitative differences
but that the difference between them diminishes close to the critical
temperature.

\begin{figure}
\psfig{file=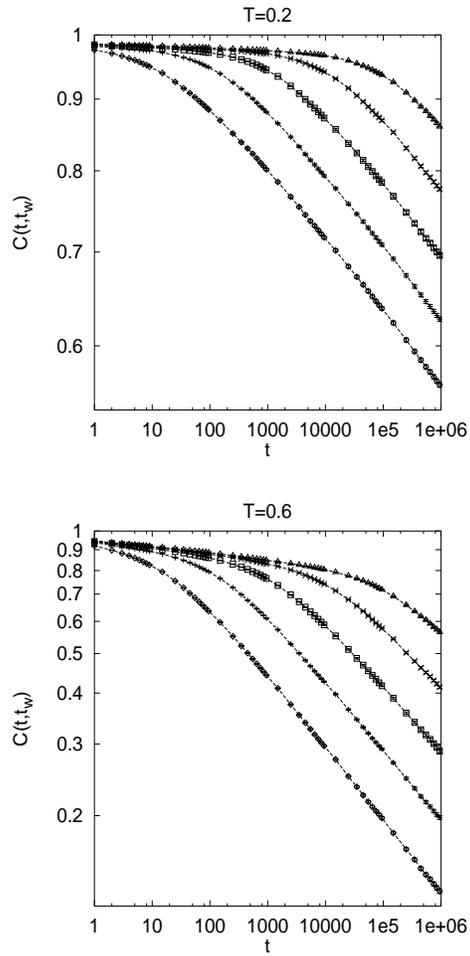,height=19cm}
\caption{The autocorrelation function $C(t,t_w)$ of the 3D EA-model for
temperatures $T=0.2$ (top) and $T=0.6$ (bottom). Note the different y-ranges
for the two temperatures. The decay of $C(t,t_w)$ is much faster at $T=0.6$.
The waiting times are $t_w=10^n \quad (n=1,\ldots,5)$ (from bottom to top).}
\label{fig3}
\end{figure}

Next we consider the autocorrelation function $C(t,t_w)$ for the waiting
times $t_w=10^n \quad (n=1,\ldots,5)$. Fig. \ref{fig3} shows
$C(t,t_w)$ for $T=0.2$ and $T=0.6$ in a log-log plot. The
curves show a characteristic crossover from a slow algebraic decay for
times $t\ll t_w$ to a faster algebraic decay for $t \gg t_w$.
This is very similar to what has been observed in other spin glass models, as
for instance in the 3D Ising spin glass with $\pm J$ couplings
\cite{rieger+-j}, a mean-field spin glass model \cite{SK} and the
one- and two-dimensional spin glass models for low temperatures
\cite{bast,kis}. The crossover from a slow quasi-equilibrium decay for
$t\ll t_w$ to a faster non-equilibrium decay for $t\gg t_w$ can be understood
in terms of equilibrated domains: During the waiting time $t_w$ the
domains have reached a certain average size. On time
scales $t\ll t_w$ processes then take place inside the
domains and thus are of a quasi-equilibrium type whereas for $t\gg t_w$
the domains continue to grow and the situation is a non-equilibrium one
resulting in a faster decay of the correlations.

The difference between the model considered
here and the models in one- and two-dimensions is that the latter do
not have a spin glass transition at non-vanishing temperature which
leads to a finite equilibration time $\tau_{eq}$ for $T>0$. Therefore
in one- and two-dimensions $C(t,t_w)$ becomes independent of $t_w$ for
$t_w > \tau_{eq}$ and the curves for different $t_w$ then coincide
\cite{kis,bast}. In three dimensions the equilibration time is
infinite below the transition temperature and thus such data collapse
is not expected to occur. This can be verified in Fig. \ref{fig1}
where the absence of data collapse of $C(t,t_w)$ even for the higher
temperature $T=0.6$ can be seen, at least on the time scales
accessible in the simulations.

It should be noted that $C(t,t_w)$ and the
waiting time dependent remanent magnetization behave similarly but
differences between them exist: Only in thermal equilibrium the
fluctuation dissipation theorem (FDT) holds and the two quantities are
simply proportional. In a non-equilibrium situation the FDT is violated
\cite{andersson,fdt} and a simple relation between them does not exist.

Algebraic fits for the long time behavior of $C(t,t_w)$
\be
C(t,t_w) \sim t^{-\lambda(T,t_w)} \qquad t \gg t_w
\label{noneqalg}
\ee
yield the set of exponents $\lambda(T,t_w)$ which are shown in Fig. \ref{fig2}.
One observes that the exponents for fixed temperature decrease with
increasing waiting time and that the waiting time dependence
is stronger for higher temperatures. The exponents for $t_w=10^4$
and $t_w=10^5$ are not shown since the available time range for the fits
is too small. We also tried to fit the data to a logarithmic function
$C(t,t_w)\sim (\ln t)^{-\lambda/\psi}$ as has been proposed in \cite{fh}
but the results were not acceptable over the whole time range.

\begin{figure}
\psfig{file=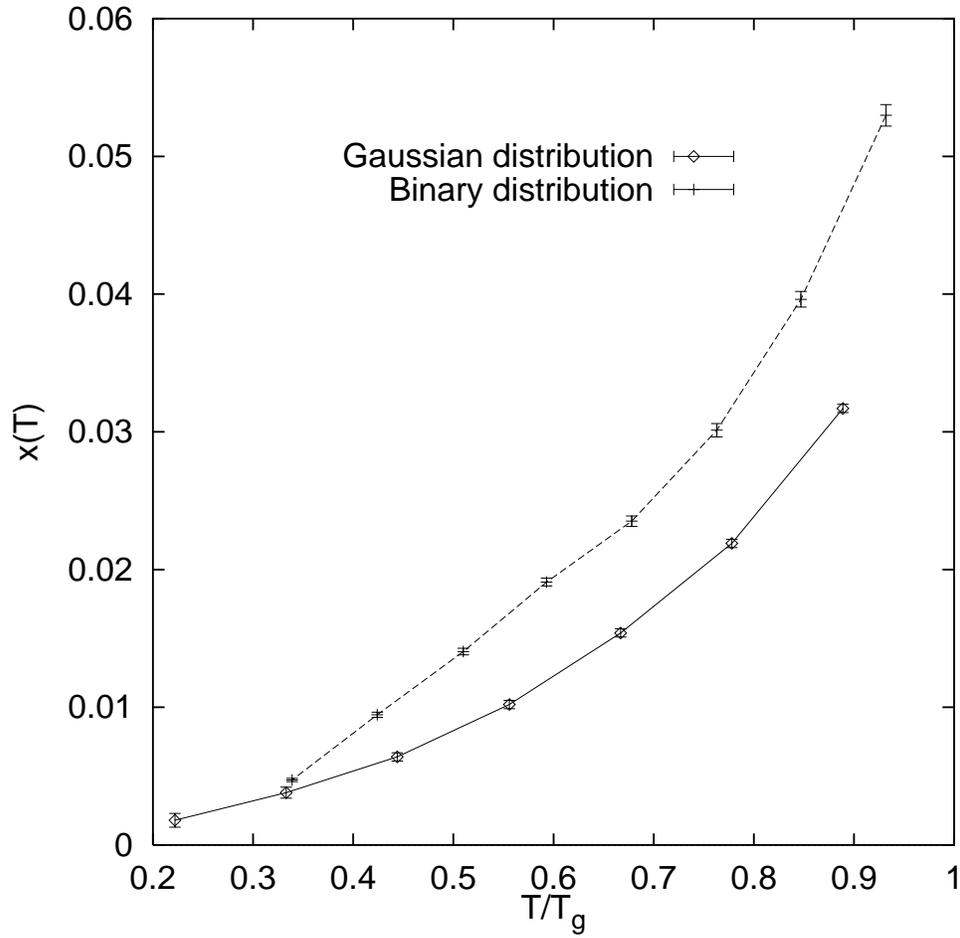,width=14cm}
\caption{The equilibrium exponent $x(T)$ of the 3D EA-model for Gaussian
and $\pm J$ couplings.}
\label{fig4}
\end{figure}

For the short time behavior $t\ll t_w$ the decay of the autocorrelation
function is also algebraic and we fitted the data to
\be
C(t,t_w)\sim t^{-x(T)} \qquad t\ll t_w \;.
\label{eqalg}
\ee
Again we tried logarithmic fits but the results for the algebraic fits
are more convincing even though it is harder to discriminate
between a logarithmic and an algebraic decay for the short time behavior
of $C(t,t_w)$ since the quasi-equilibrium exponent $x(T)$ is much smaller than
the non-equilibrium exponent $\lambda(T,t_w)$. As can already be seen in Fig.
\ref{fig3}
the exponent $x(T)$ for fixed temperature is independent of the waiting
time and thus the data for $t_w=10^5$ have been used for the fits
since the quasi-equilibrium time range is longest there.
$x(T)$ is shown in Fig. \ref{fig4}. For comparison the corresponding
quantities for the 3D spin glass model with a binary distribution of
the couplings which were determined in \cite{rieger+-j} are also shown.
One observes that $x(T)$ increases with temperature and that the
shape of the curves is similar for both distributions but
quantitative differences between the exponents exist which however depend
on the assumed critical temperatures. For the model with Gaussian
couplings $T_g$ is not known to great accuracy. To our knowledge three
different values for $T_g$ have been determined:
$T_g=0.9-1.0$ \cite{bhatttc}, $T_g=1.0\pm 0.2$ \cite{mcmillan} and
$T_g=0.8\pm 0.1$ \cite{braytc}. If one assumes $T_g=1.0$ instead of
$T_g=0.9$, the curve for Gaussian couplings in Fig. \ref{fig4} is shifted
to the left and both curves then roughly coincide. The exponent $x(T)$
has also been measured experimentally in the short range Ising spin glass
$\rm Fe_{0.5}Mn_{0.5}TiO_3$ \cite{gunnarsson} and close to $T_g$
($T/T_g=1.029$) $x=0.07$ was found.

\begin{figure}
\psfig{file=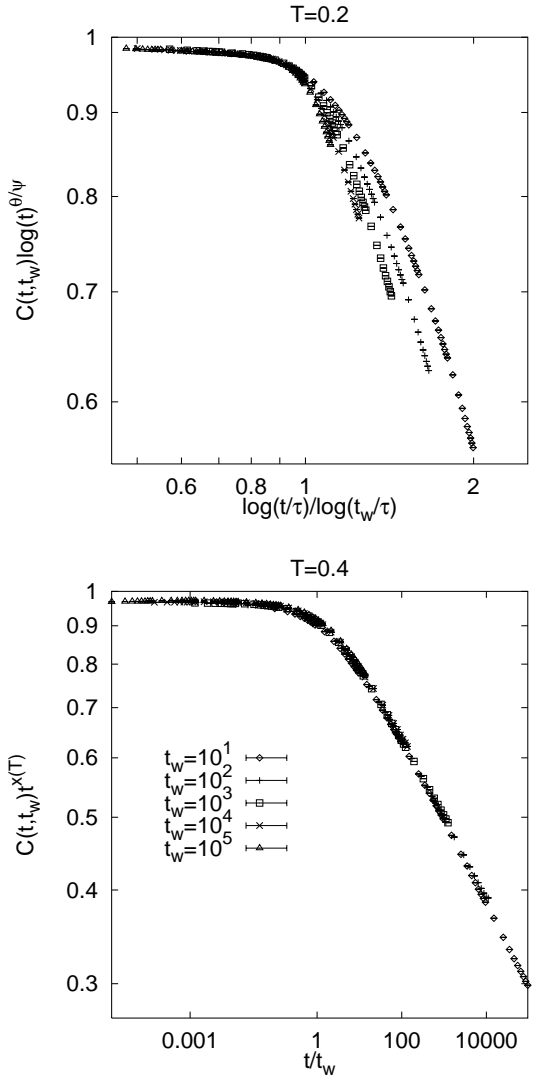,width=11.5cm}
\caption{Scaling of the autocorrelation function $C(t,t_w)$ of the 3D
EA-model. Logarithmic scaling plot (top) and $t/t_w$ scaling (bottom).
Obviously $C(t,t_w)$ scales with $t/t_w$.}
\label{fig5}
\end{figure}

Next we examine the scaling behavior of $C(t,t_w)$. The droplet
theory of Fisher and Huse \cite{fh} predicts a scaling $C(t,t_w) \sim
\tilde{\Phi}(\ln t_w / \ln t )$ whereas in MC simulations of
some of the above mentioned spin glass models \cite{rieger+-j,kis,bast}
and in a mean-field like model \cite{SK} a scaling law
$C(t,t_w) \sim \Phi(t/t_w)$
was found. In the present model a logarithmic scaling
can be clearly ruled out whereas the data do speak in favour of a
$t/t_w$ scaling. This can be seen from Fig. \ref{fig5} where
scaling plots of the form
\begin{equation}
  C(t,t_w)\sim (\ln t)^{-\theta / \psi} \tilde{\Phi}\left( \frac{\ln
      (t/ \tau)}{\ln (t_w / \tau)} \right)
\label{logscale}
\end{equation}
and
\begin{equation}
  C(t,t_w)=c_Tt^{-x(T)}\Phi_T(t/t_w)\; .
\label{algscale}
\end{equation}
are shown. $x(T)$ is the exponent describing the quasi-equilibrium decay
of $C(t,t_w)$ in (\ref{eqalg}).
For the logarithmic scaling law the ratio $\theta / \psi$
and the variable $\tau$ have been used as parameters in order to
obtain maximum data collapse. Best results were obtained for $\theta /
\psi =0$ but scaling remains unsatisfactory which is similar to what
is reported in \cite{bouchaud} where the scaling behavior of experimental
data for the waiting time dependent remanent magnetization is investigated.

\section{Domain growth in the 3D spin glass model}
\label{sectionspatial3d}

Since it has been proposed that an extremely slow domain growth is the
reason why aging can be observed in short range spin glasses
\cite{fh,koper} and because domain growth laws are an important
ingredience of phenomenological models as pointed out in the introduction
we also examined spatial correlations in the simulations
in order to calculate a time dependent correlation length. This
correlation length can be thought of as a measure of the average
domain size in the system after the temperature quench.  The situation
is similar to ferromagnets, where the non-equilibrium dynamics after a
temperature quench is characterized by growth of domains where the
spins are aligned as in either one of the two ground states.

In spin glasses the identification of domains is more difficult since the
ground state is unknown in three dimensions.  A suitable correlation
function has to be used instead. The generalization of the usual
ferromagnetic correlation function $G(r)=\langle S_i S_{i+r} \rangle$ to spin
glasses is $G(r)=[\langle S_i S_{i+r} \rangle^2]_{av}$. However, as became
obvious in MC simulations of the two-dimensional spin glass
\cite{bast}, the square makes it difficult to get good statistics for
the quantity $G(r)$ since it leads to a positive bias in the signal.
Instead, a replica method has been used where the square is
substituted by the spins of two replicas of the system, i.e. systems
with identical couplings $J_{ij}$ but different initial conditions and
thermal noise. This leads to the generalized time dependent
correlation function
\begin{equation}
  G_T(r,t_w)=\frac{1}{N}\sum_{i=1}^N \frac{1}{t_w} \sum_{t=t_w}^{2t_w-1}
  [\langle S_i^a(t)S_{i+r}^a(t)S_i^b(t)S_{i+r}^b(t) \rangle]_{av}
\label{spatialcorr}
\end{equation}
which is suitable to measure the expected domain growth. Note that in
two dimensions the (up to a global spin flip) unique ground state for
Gaussian couplings can be calculated and be used in (\ref{spatialcorr})
instead of the spins of one of the replicas. The results of such an
analysis are presented in Sec. \ref{sectionspatial2d}.

The waiting times were chosen to be $t_w=4^n \quad (n=1,\ldots,10)$ or
$t_w=5^n \quad (n=1,\ldots,8)$ respectively. To improve the
statistics of $G_T(r,t_w)$ for smaller waiting times the number of
samples was chosen to be waiting time dependent such that the product
$\#\mbox{samples}\cdot t_w$ was approximately constant.  For larger
waiting times the average was taken over at least 128 samples; for the
smallest waiting time ($t_w=4$) 520000 samples have been used.

\begin{figure}
\psfig{file=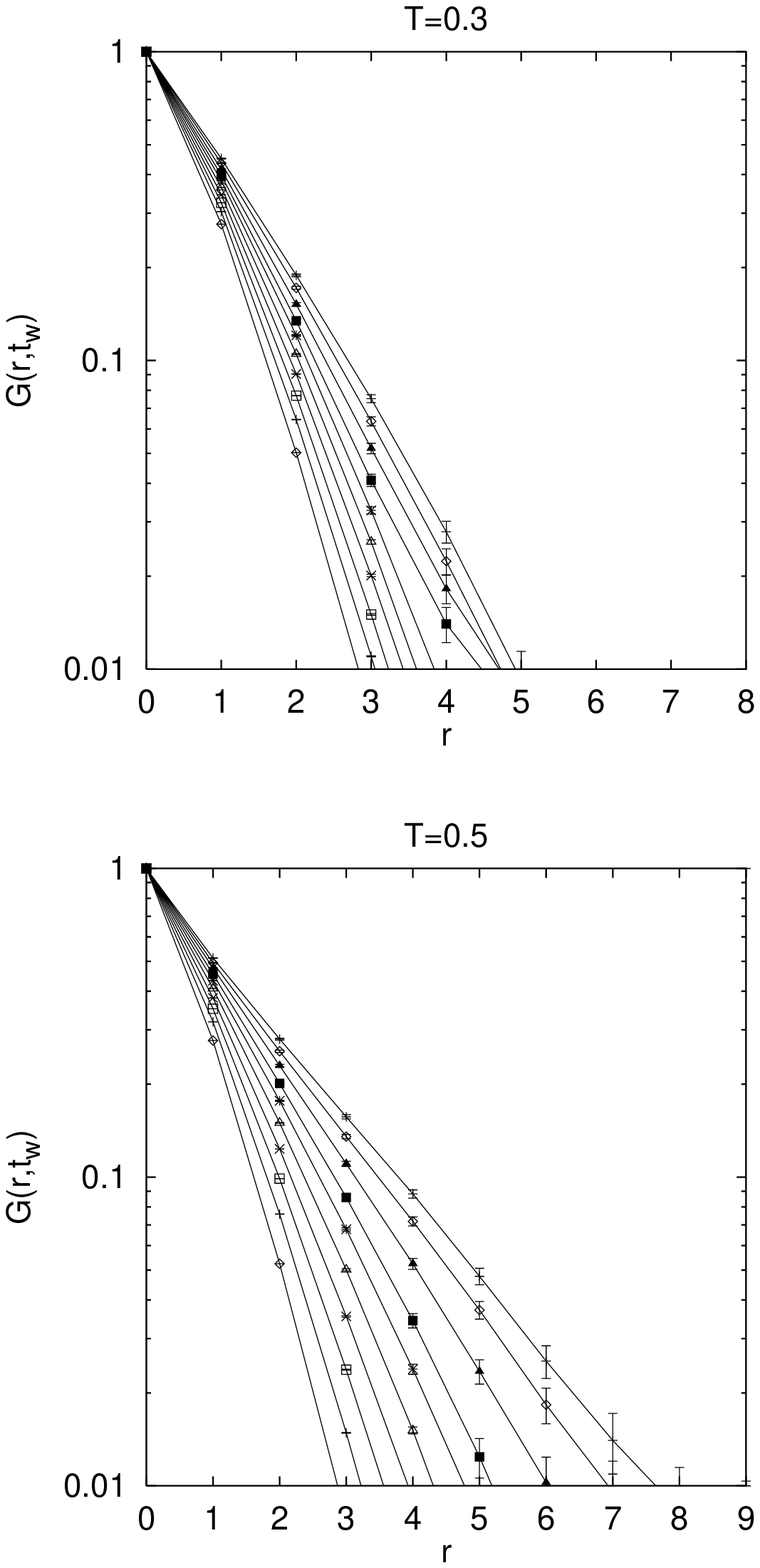,width=11.5cm}
\caption{The spatial correlation function $G(r,t_w)$ of the 3D EA-model
for temperatures $T=0.4$ and $T=0.5$. The waiting times are $t_w=4^n \quad
\mbox{(n=1,\ldots,10)}$ (from bottom to top). }
\label{fig6}
\end{figure}

$G_T(r,t_w)$ is shown for two different temperatures in a log-linear
plot in Fig. \ref{fig6}.  The different curves correspond to different
waiting times $t_w$. One observes that the correlations fall off
rapidly with $r$ for small waiting times which means that the typical
domain size is only of the order of a few lattice spacings. For larger
waiting times, however, the correlations increase. This is caused by
the growth of ordered domains as mentioned earlier. In the log-linear
plot the curves look approximately linear which means that the decay of
$G_T(r,t_w)$ is roughly exponential. In contrast to the two-dimensional
spin glass the curves for larger waiting times do not coincide which means that
the system is not equilibrated even after the longest waiting time in
the simulations. This is in agreement with what has been said about
the autocorrelation function $C(t,t_w)$ in the previous section.

In principle an effective correlation length can be determined from
$G_T(r,t_w)$ by calculating the integral
\begin{equation}
  \xi_T(t_w)=2\int_0^{\infty}G_T(r,t_w)\;dr\;.
\label{xiint}
\end{equation}
This definition is motivated by the fact that for a purely exponential
decay $G_T(r,t_w) \sim \exp(-r/\xi)$ this effective correlation length
is equal to the length scale $\xi$. The factor 2 in (\ref{xiint}) is
introduced since $G(r,t_w)$ in (\ref{spatialcorr}) measures the square
of a correlation function.

\begin{figure}
\psfig{file=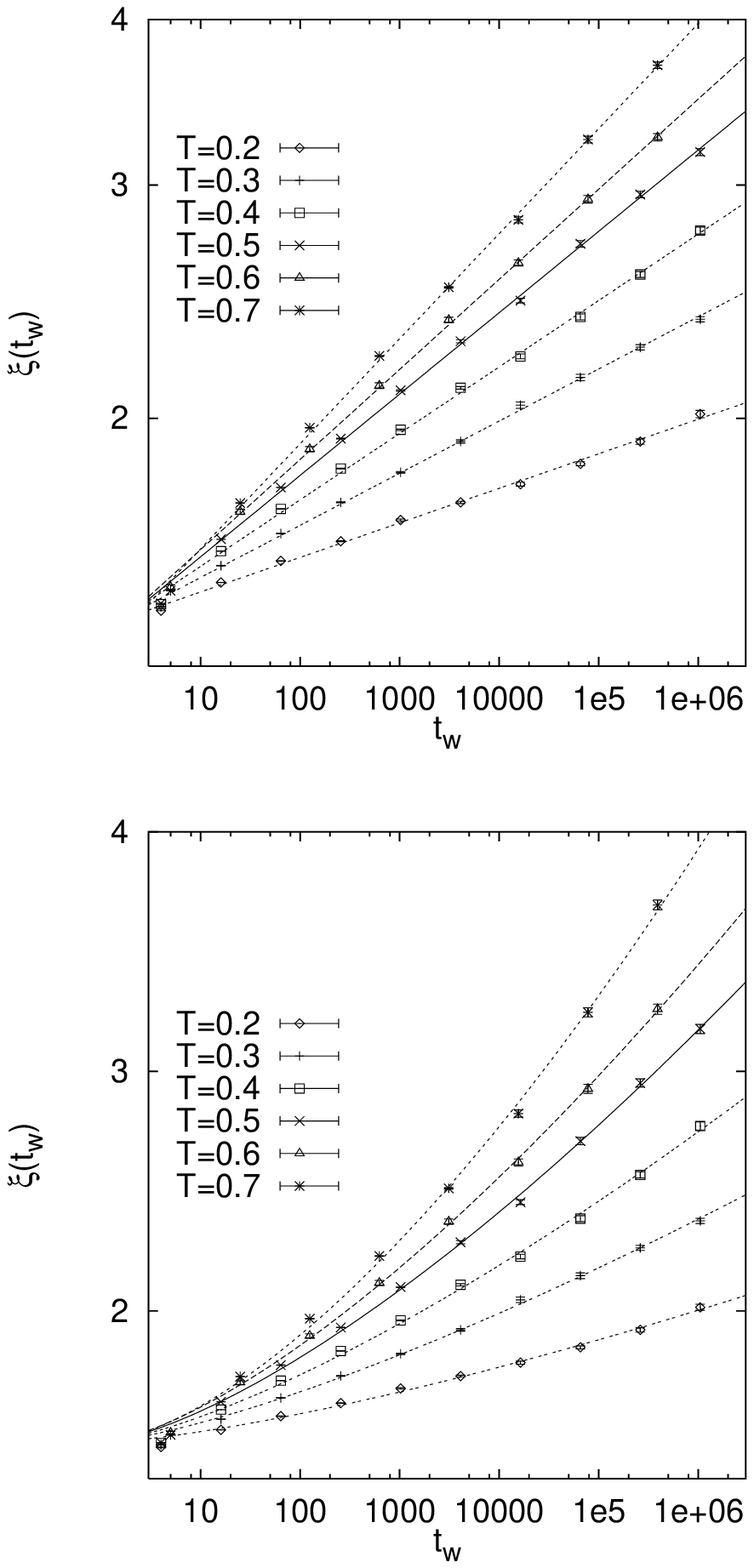,height=18cm}
\caption{Algebraic (top) and logarithmic (bottom) fits of the correlation
length $\xi(t_w)$ of the 3D EA-model for different temperatures.}
\label{fig7}
\end{figure}

When evaluating the integral
(\ref{xiint}) the periodic boundary conditions have to be taken into
account. In a system of length $L$ the correlations can only be
calculated up to $r=L/2$. Furthermore, for fixed $r$ one actually
measures the correlation $G(r,t_w)+G(L-r,t_w)$.  Thus the integral $I$
over the function measured in the simulations has contributions from
$G(r,t_w)$ and $G(L-r,t_w)$:
\be
I=\int_0^{L/2}G(r,t_w)\;dr + \int_0^{L/2}G(L-r,t_w)\; dr \; .
\ee
Assuming an exponential decay of $G(r,t_w)\sim \exp(-r/\xi(t_w))$
one obtains an implicit equation for $\xi(t_w)$:
\be I=\xi(t_w)(1-\exp(-L/\xi(t_w)))\;.  \ee
The resulting values of $\xi(t_w)$ are shown for different
temperatures in Fig. \ref{fig7}. On the left hand side logarithmic
fits
\be \xi(t_w) - \xi_0 \sim (\log(t_w))^{1/\psi}
\label{loggrow}
\ee
and on the right hand side algebraic fits
\be
\xi(t_w)\sim t_w^{\alpha(T)}
\label{alggrow}
\ee
are plotted. As can be seen, both fits are of the same quality
and in terms of a $\chi^2$-test there is no difference between them.
Via the algebraic fit one obtains a set of temperature dependent
exponents $\alpha(T)$ which increase approximately linearly with
temperature, for instance it is $\alpha(T=0.2)=0.026$ and
$\alpha(T=0.7)=0.081$. These exponents are very small since the
correlation length grows only slowly with time. Even for the highest
temperature $T=0.7$ used in the simulations $\xi(t_w)$ is less than
four lattice spacings after $10^6$ MCS. The logarithmic fit
yields a roughly temperature independent value of $\psi\approx 0.71\pm 0.02$
which is in agreement with the bound $\psi \le d-1$ given in \cite{fh}.
Recently a value of $\psi \approx 0.8$ has been determined experimentally
\cite{nophase} in the Ising spin glass $\rm Fe_{0.5}Mn_{0.5}TiO_3$
via dynamic scaling.

\begin{figure}
\psfig{file=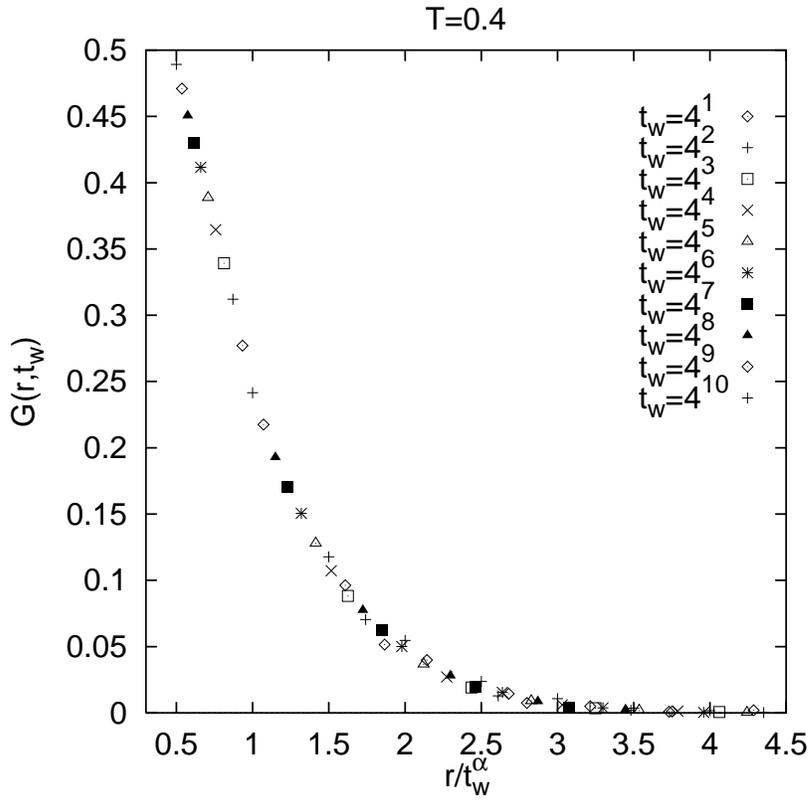,width=14cm}
\caption{Scaling plot of $G(r,t_w)$ of the 3D EA-model for $T=0.4$. The
characteristic length scale $\xi(t_w)$ grows as $\xi(t_w) \sim
t_w^{\tilde{\alpha}(T)}$}
\label{fig8}
\end{figure}

Considering the scaling law (\ref{algscale}) of the autocorrelation function
one expects a similar scaling law to hold for
the spatial correlation function $G(r,t_w)$. This can be seen in
Fig. \ref{fig8} where a scaling plot of the form
\be
G(r,t_w) \sim \tilde{g}(r/\tilde{\xi}(t_w))
\label{spatialscale}
\ee
with the characteristic length scale
$\tilde{\xi}(t_w) \sim t_w^{\tilde{\alpha}(T)}$ is shown. The exponents
$\tilde{\alpha}(T)$ have been determined such
that best scaling behavior was achieved. They concur within the
error bars with the exponents $\alpha(T)$ determined by the above
integration procedure. Thus the characteristic length scale $\xi(t_w)$
can be obtained via the scaling law (\ref{spatialscale}) in a different
and more simple way than by (\ref{xiint}) since the problems arising
from the periodic boundary conditions and the extrapolation to
infinity are avoided. The numerical errors are roughly the same for both
methods.

Considering the algebraic decay (\ref{noneqalg}), (\ref{eqalg}) and
the scaling behavior (\ref{algscale}) of the autocorrelation function
$C(t,t_w)$ an algebraic growth law (\ref{alggrow}) for the correlation
length gives a more consistent picture from our point of view for the
following reason: Assuming a logarithmic growth law (\ref{loggrow})
one obtains a logarithmic decay of $C(t,t_w)$ since
$C(t,t_w)\sim [L(t_w)/L(t+t_w)]^{\lambda}$ and thus also logarithmic scaling
which is not compatible with the results of the simulations
as was shown above. However, with an algebraic growth law
(\ref{alggrow}) the autocorrelation function decays algebraically and
scales with $t/t_w$ which is exactly what we found.
A logarithmic growth law
was used in \cite{fh} by Fisher and Huse. They assumed a particular
scaling law $B(L)\sim L^{\psi}$ for the (free-) energy barriers on
length scale $L$ and via activated dynamics obtained $R(t_w) \sim
(\ln t)^{1/\psi}$ for the average domain size. As has already been noted in
\cite{rieger+-j} a modified
activated dynamics scenario with a scaling law
\be
B(L) \sim \Delta (T)\log(L)
\label{logbarrier}
\ee
for the energy barriers leads to algebraic domain growth
$\xi(t_w)\sim t_w^{\alpha(T)}$ with $\alpha(T)=T/\Delta(T)$ and thus is
appropriate to describe the results of the simulations. In
\cite{annealing} the dependence of the barrier height on
length scale $L$ has been examined explicitly by an annealing procedure
and there was no evidence for an algebraic dependence but also
a logarithmic law $B_L \sim \log L$ was found thus supporting
our results. The predictions for the different scaling assumptions are
summarized in Table \ref{tab1}.

Finally we want discuss the concept of an overlap length. It has been
argued \cite{fh,koper,overlap} that the correlations in the spin glass phase
are extremely sensitive to temperature- and field changes. This sensitivity
should be observable via the correlation function
\be
\Theta(r,T_1,T_2)=[\langle S_iS_{i+r}\rangle_{T_1}
                  \langle S_iS_{i+r}\rangle_{T_2}]_{av} \;.
\label{theta}
\ee
This function is expected to decay as
$\Theta(r,T_1,T_2) \sim \exp(-r/L(T_1,T_2))$ with the length
scale $L(T_1,T_2)$ being the overlap length. Thus the correlations
at temperatures $T_1$ and $T_2$ are the
same for length scales smaller than $L(T_1,T_2)$ whereas the
correlations on larger length scales are completely destroyed.

It is expected \cite{overlap} that $L(T_1,T_2)$ is finite even below
the spin glass transition temperature where the usual correlation length
$\xi_{eq}$ is infinite. In thermal equilibrium
$L(T_1,T_2)\sim |T_1-T_2|^{-1/\zeta}$ with {\em positive}
$\zeta=d_s/2-\theta$ is predicted \cite{overlap}. $d_s$ is the
fractal dimension of the droplets in the droplet theory and $\theta$
the exponent describing the free energy of the droplets $F_L\sim L^{\theta}$.

\begin{figure}
\psfig{file=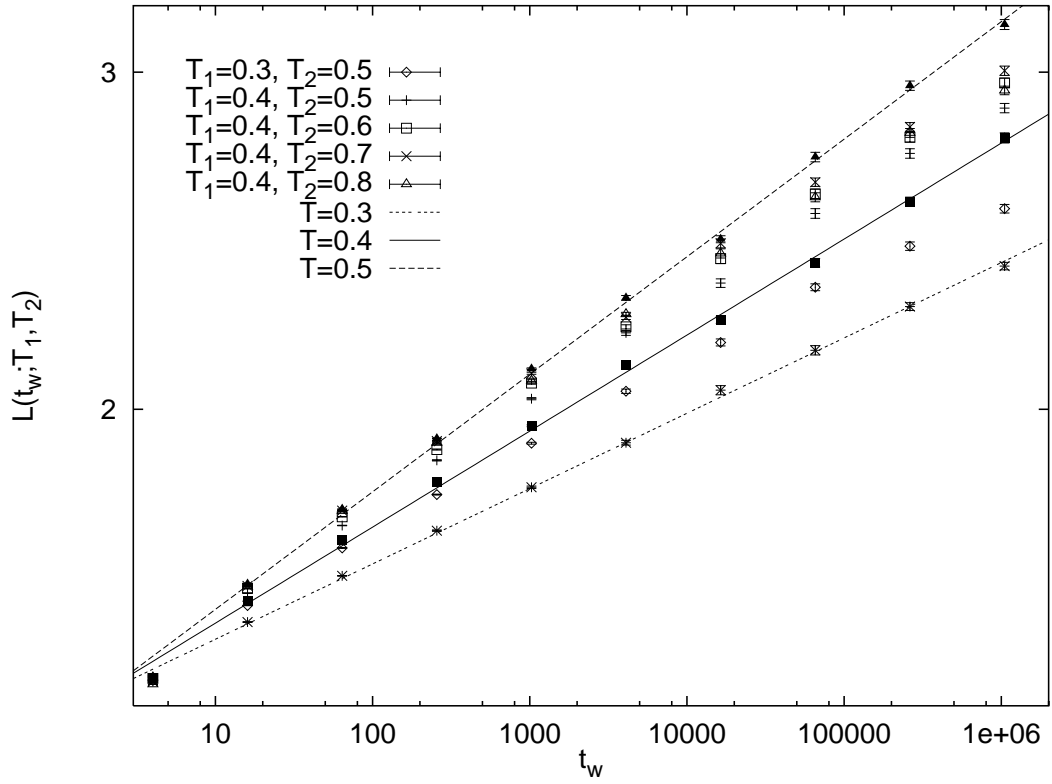,width=14cm}
\caption{The overlap length $L(t_w;T_1,T_2)$ of the 3D EA-model and the
correlation length $\xi(t_w)$ for different temperatures.}
\label{fig9}
\end{figure}

To check the existence of such an overlap length we calculated the
(non-equilibrium) correlation function
(\ref{spatialcorr}) but with the two replicas $a$ and $b$ having the
temperatures $T_1$ and $T_2$ respectively, which yields
$\Theta(r,t_w;T_1,T_2)$. Note that $\lim_{t_w\rightarrow\infty}
\Theta(r,t_w;T_1,T_2) = \Theta(r,T_1,T_2)$ as defined in (\ref{theta}).
The resulting curves
look very similar to the correlation functions in Fig. \ref{fig6}.
Thus we determined the waiting time dependent overlap length in the same way
as the correlation length
$\xi(t_w)$ by calculating the integral over the function
$\Theta(r,t_w,T_1,T_2)$ as in (\ref{xiint}):
\be
L(t_w;T_1,T_2) = 2\int_0^{\infty} \Theta(r,t_w;T_1,T_2) \; dr \; .
\ee
The results are shown in Fig.
\ref{fig9} where $L(t_w;T_1,T_2)$ and for comparison the correlation length
$\xi_T(t_w)$ are shown. One observes that the overlap length for
temperatures $T_1$ and $T_2$ $(T_1 < T_2)$ increases faster than
the correlation length with both temperatures set to $T_1$ as can be
seen for example by comparing the overlap length for $T_1=0.4$ and
$T_2=0.6$ with the correlation length for $T=0.4$. With increasing
temperature difference $T_2-T_1$ the overlap length does not change
significantly (see the curves for $T_1=0.4$ and $T_2$=0.5, 0.6, 0.7 and $0.8$
respectively), in particular it does not decrease with increasing
temperature difference, contrary to what one might expect according
to the arguments given above. This means that either an overlap length in
its original sense does not exist or more likely, that the overlap
length is much larger than the correlation lengths reached in the simulations,
which would make it impossible to observe the effects of its existence on the
accessible time scales.
The only effect of setting one replica to $T_1$ and
the other replica to $T_2$ $(T_1 < T_2)$ is that the faster dynamics
at the higher temperature $T_2$ leads to faster domain growth
and thus the correlations increase faster than with both
replicas having temperature $T_1$. However, the increase of the
correlations seems to be limited by the lower of the two temperatures.

This interpretation is compatible with the outcome of temperature cycling
experiments for the same model (with binary couplings) \cite{tcycle}.
However, note that similar to the situation with regards to experiments
\cite{lefloch,refregier,vincent,granberg2,mattson}, also other
interpretations have been suggested \cite{andersson2}.

\section{Domain growth in the 2D spin glass model}
\label{sectionspatial2d}
In this section we consider the two-dimensional spin glass model and
present the results of an alternative method to identify domains.
The model is the two-dimensional analogue to the 3D-model introduced in
Sec. \ref{sectionmodel}, i.e. we have nearest-neighbour interactions,
periodic boundary conditions, Gaussian couplings and Glauber dynamics.

In contrast to the 3D-model (\ref{hami}), it is now possible
to calculate the ground state of the model in two dimensions.
A very fast implementation of a branch and cut algorithm \cite{br-cut} has
been used making it possible to obtain ground states for lattice
sizes up to $150 \times 150$ on an ordinary workstation. Thus
it is not necessary to introduce a replica system to identify domains
since the ground state can be used instead. An analysis using a
replica system has been performed in a previous work \cite{bast} and
we compare the results of both methods.

\begin{figure}
\psfig{file=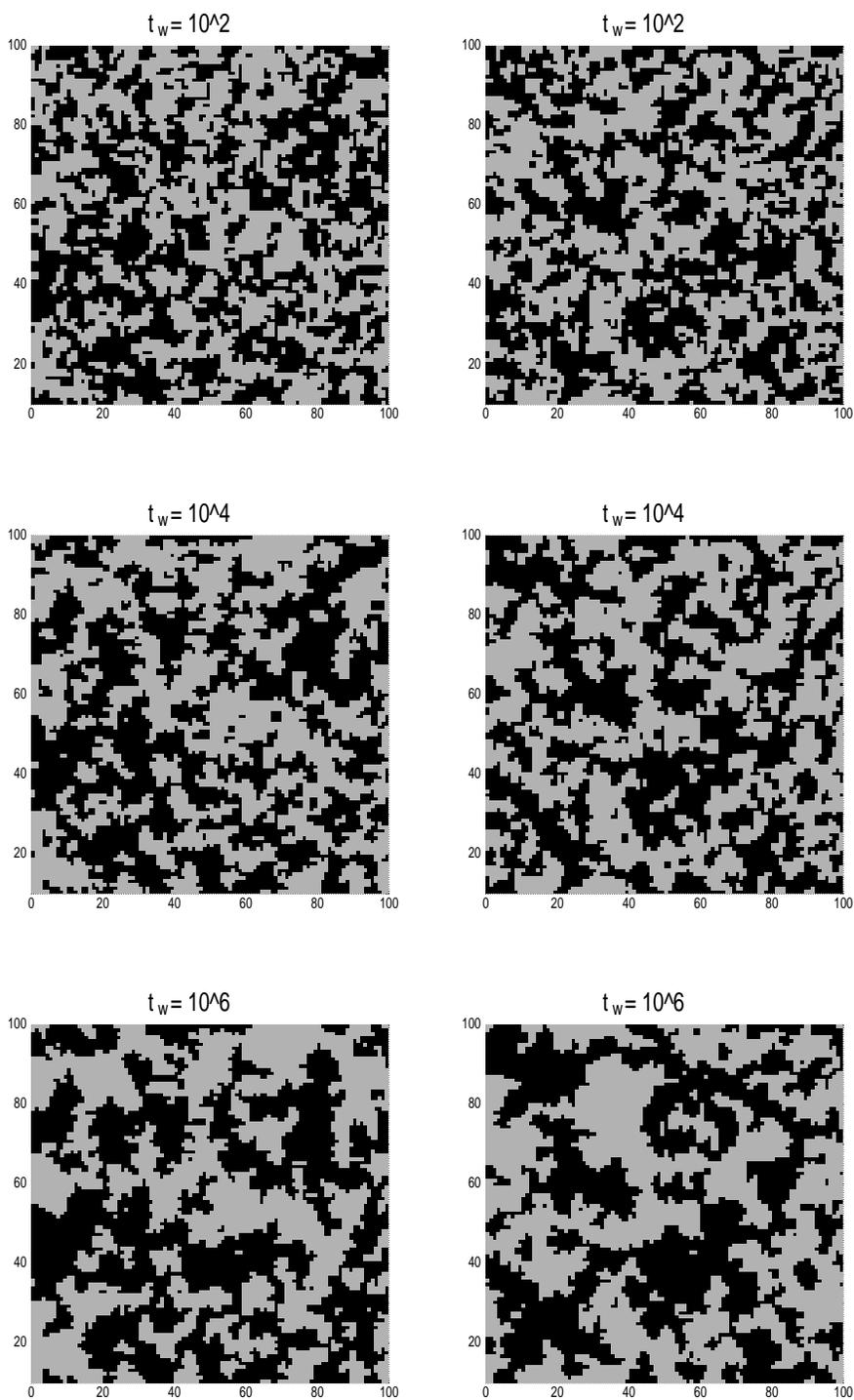,width=13cm}
\caption{Domain growth in the 2D EA-model with Gaussian couplings. The
right hand side shows the domains relative to the ground state and the left
hand side shows the same configuration relative to a replica system.}
\label{fig10}
\end{figure}

In Fig.\ref{fig10} the time evolution of an initially random spin
configuration is shown. The domains have been identified by
calculating for each spin $i$
\be
q_i^{EA}(t_w) = \frac{1}{\Delta}\sum_{t=t_w}^{\Delta -1} S_i^a(t)S_i^b(t)
\ee
for the replica method and
\be
q_i^{gs}(t_w)=\frac{1}{\Delta}\sum_{t=t_w}^{\Delta-1} S_i(t)S_i^0
\ee
by using the ground state. $\Delta$ is a suitably chosen time window
and $S_i^0$ denotes the ground state of the spin at site $i$.
In both methods the same couplings were used. Obviously both methods
show an increasing average domain size but the method using ground states
shows larger domains since in the replica method one has two thermally
active systems. Note that the domain structures look very different in both
methods even though the initial configuration of one of the replicas
was the same as that of the system used in the ground state method.
In contrast to ferromagnetic systems even for very large waiting times
very small domains exist. These are either very stable clusters because
strong bonds have to be broken to flip the spins or new domains within
the bigger ones
appear since less strongly bound spins initialize the formation of a new
domain.

To compare both methods quantitatively we measured the correlation function
\begin{equation}
   G(r,t_w) = \frac{1}{N}\sum\limits_{i=1}^N\frac{1}{t_w}
   \sum\limits_{t=t_w}^{2t_w-1}[\langle S_i(t)S_{i+r}(t)S_i^0S_{i+r}^0
    \rangle ]_{av}
\label{spatialcorrgrnd}
\end{equation}
in the simulations.
This correlation function is different from the one
(\ref{spatialcorr}) used in the 3D-model in that the spins of the replica
system have been substituted by the ground state. We used 128
ground states of $40\times 40$ lattices to perform our investigation. In order
to obtain good statistics also for smaller waiting times we averaged over up
to 4000 different initial configurations and realizations of the
thermal noise for each ground state.

\begin{figure}
\psfig{file=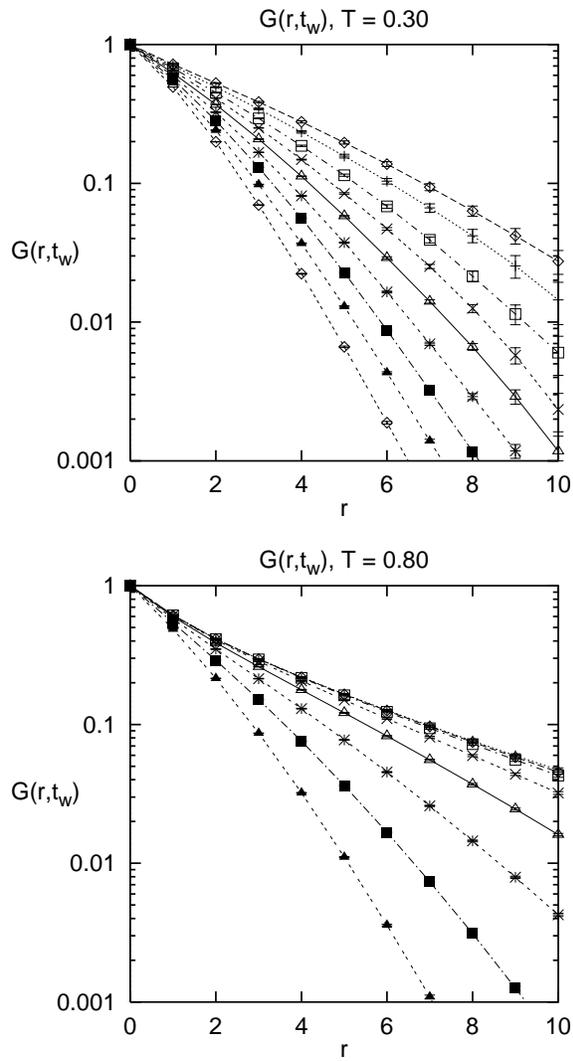,width=12cm}
\caption{The spatial correlation function $G(r,t_w)$ of the 2D EA-model
spin glass and temperatures $T=0.3$ and $T=0.8$. The waiting times are $t_w=5^n
\quad \mbox{(n=1,\ldots,9)}$. }
\label{fig11}
\end{figure}

Fig.\ref{fig11} shows $G(r,t_w)$ for two different temperatures in a log-linear
plot. As in the 3D-model $G(r,t_w)$ decreases rapidly with $r$
but the decay is slower than in the 3D-model. For lower temperatures
(see the curve for $T=0.3$) $G(r,t_w)$ increases monotonously
with waiting time as in the 3D-model whereas for higher temperatures
(see $G(r,t_w)$ for $T=0.8$) the curves for higher waiting times coincide.
This means that the system is equilibrated at a certain waiting time
and the correlations take on their equilibrium values which
is due to the fact that the 2D-model has $T_g=0$ and thus the
equilibration time is finite. For lower temperatures the equilibration
time is larger than the simulation time which makes the resulting curves
qualitatively indistinguishable from the ones in the 3D-model.

\begin{figure}
\psfig{file=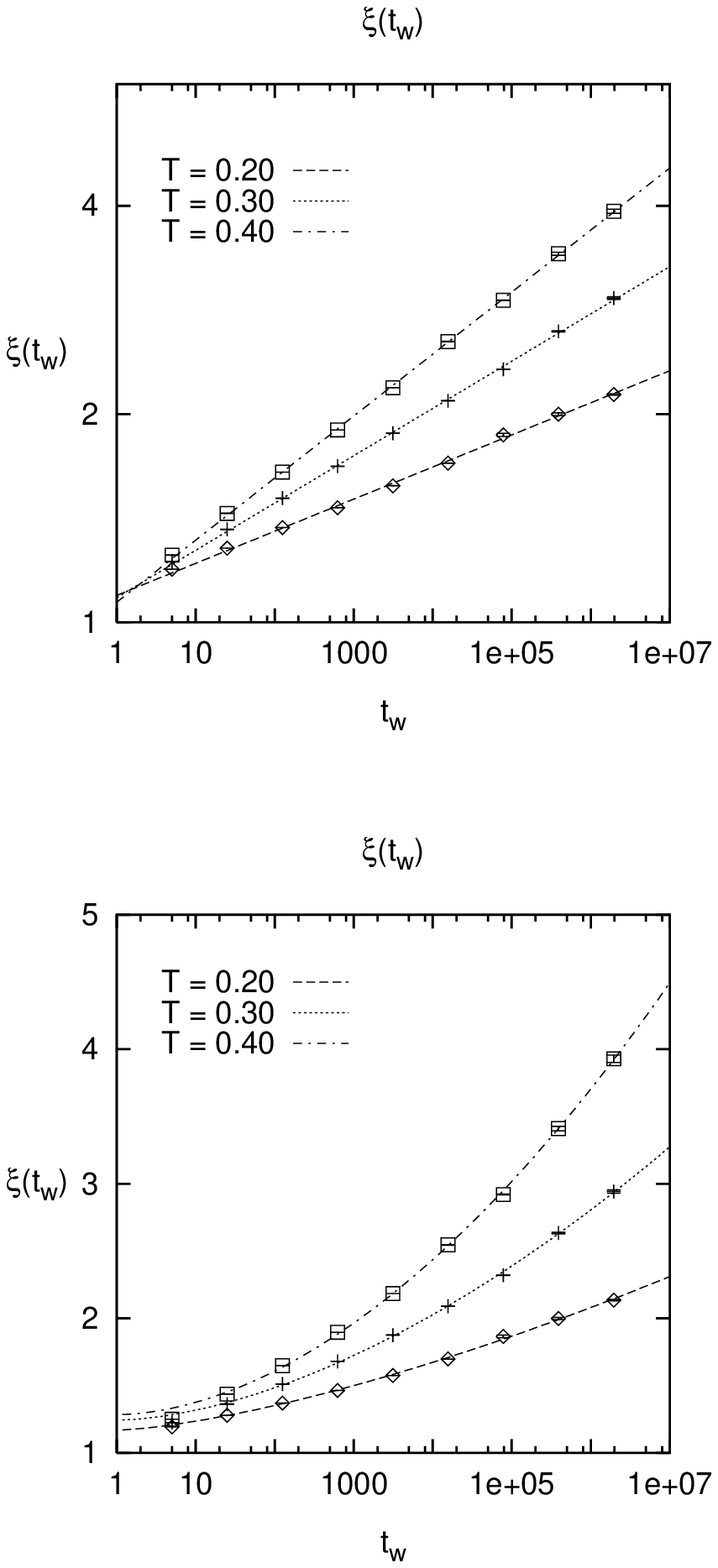,width=14cm}
\caption{Algebraic (top) and logarithmic (bottom) fits of the correlation
length $\xi(t_w)$ of the 2D EA-model for different temperatures.}
\label{fig12}
\end{figure}

\begin{figure}
\psfig{file=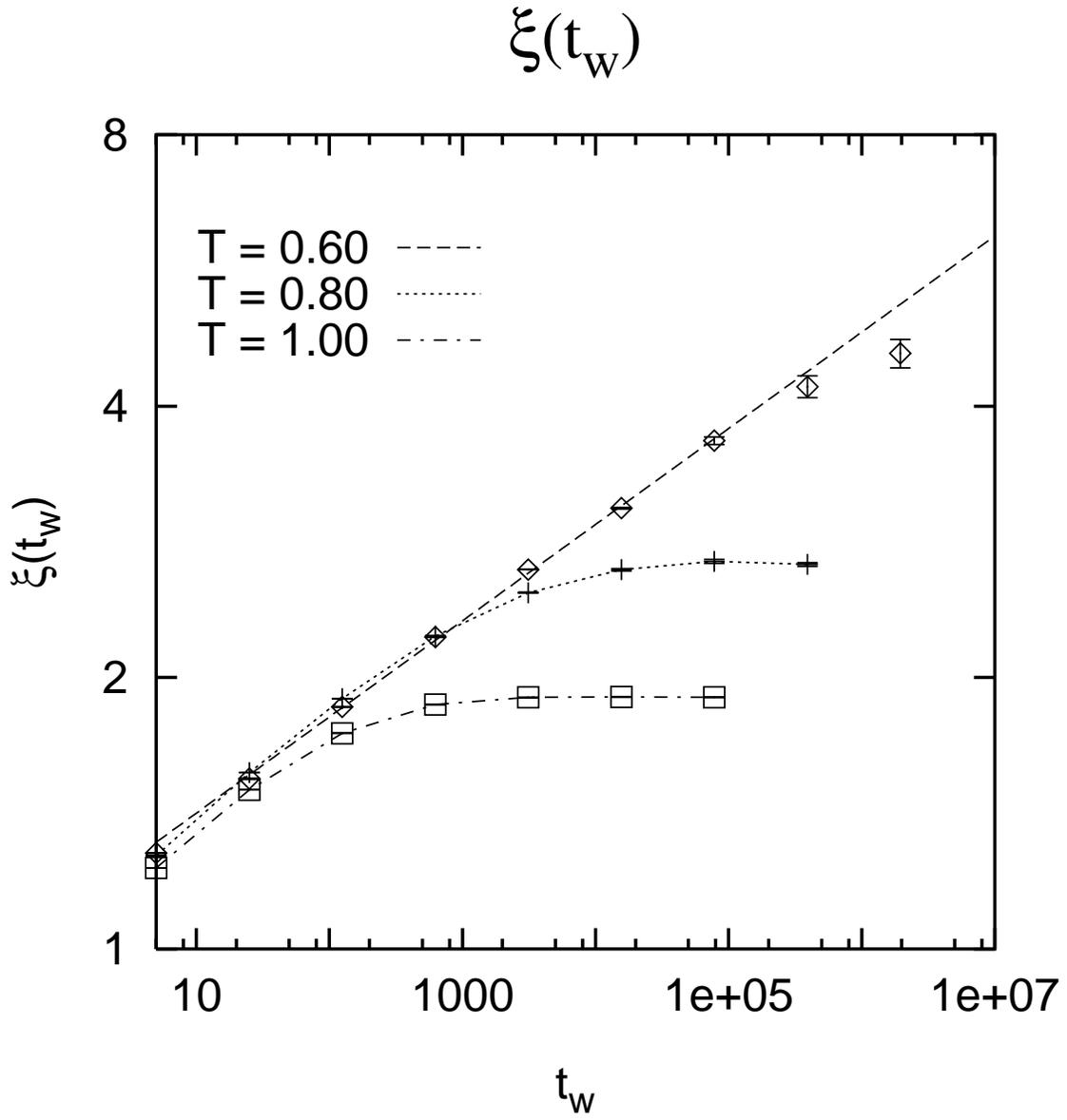,width=14cm}
\caption{The correlation length $\xi(t_w)$ of the 2D EA-model for higher
temperatures.}
\label{fig13}
\end{figure}

As in the 3D-model an effective correlation length can be calculated from
$G(r,t_w)$ by
\begin{equation}
  \xi(t_w)= \int\limits^\infty_0 G(r,t_w)\; dr\;.
 \label{xiintgrnd}
\end{equation}
The factor 2 present in (\ref{xiint}) is left out here because the
correlation function defined in
(\ref{spatialcorrgrnd}) is similar to a ferromagnetic correlation
function. Our results for $\xi(t_w)$ are shown
in Fig.\ref{fig12} and for somewhat higher temperatures in Fig. \ref{fig13}.
Again logarithmic fits according to
(\ref{loggrow}) and algebraic fits according to (\ref{alggrow}) have been done.
Only the low temperatures have been considered since for higher temperatures
the correlation length saturates for small waiting times thus limiting
the time range for the fits. Obviously our data can nicely be fitted
to the logarithmic growth law as well as to the algebraic law. Comparing the
values of the correlation length to the earlier investigation \cite{bast}
where the replica method has been used, we obtain smaller values for the
correlation length. However, the exponents $\alpha(T=0.2)=0.046$,
$\alpha(T=0.3)=0.068$ and $\alpha(T=0.4)=0.090$ describing the algebraic
growth of the correlation length and also the roughly
temperature independent value of $\psi = 0.61\pm0.05$ agree within the
errorbars with those obtained from the replica method.
Furthermore it should be mentioned that the exponents $\alpha(T)$ grow
linearly with temperature.

\begin{figure}
\psfig{file=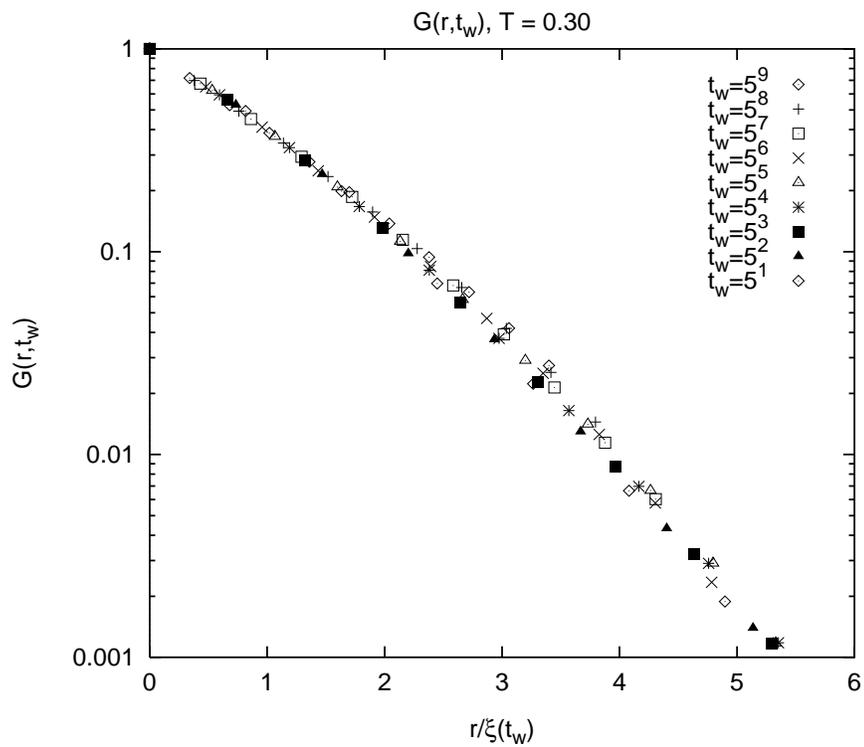,width=14cm}
\caption{Scaling plot of $G(r,t_w)$ of the 2d EA-model for $T=0.3$}
\label{fig14}
\end{figure}

The scaling behavior of $G(r,t_w)$ is shown in Fig. \ref{fig14}. As in
(\ref{spatialscale}) for the 3D-model the exponents $\tilde{\alpha}(T)$
for the characteristic length scale $\xi(t_w)\sim t_w^{\tilde{\alpha}(T)}$
have been determined such that best scaling was achieved. Analogous to the
3D-model the values for the resulting exponents $\tilde{\alpha}(T)$
also agree within the errorbars with those obtained from the integration
method.

\section{Conclusion}
\label{sectionconclusion}
Concluding we examined the non-equilibrium dynamics of the two- and three-
dimensional spin glass model in detail by calculating correlation functions in
time and space. In the 3D-model algebraic decay of the autocorrelation
function including the remanent magnetization was found. The exponents
of the decay of the remanent magnetization are in quantitative agreement
with experimental values. The autocorrelation function was found to scale
with $t/t_w$ and a typical aging scenario was established.
A comparison of the exponents describing the quasi-equilibrium decay of
$C(t,t_w)$ for a Gaussian and a binary distribution of the couplings showed
that they take on roughly the same values but it is hard to decide
whether they are universal quantities or not since the critical temperature
for the model with Gaussian couplings is not known to adequate accuracy.
The investigation of the spatial correlations showed that the average domain
size depends algebraically on the waiting time. These results can explained
consistently within a modified droplet theory where the original barrier
law $B(L)\sim L^{\psi}$ is replaced by $B(L)\sim \log L$. The investigation
of the sensitivity of the spatial correlations with respect to temperature
changes did not show evidence for the existence of a finite overlap length on
time scales accessible in the simulations.

The investigation of the spatial correlations in the two dimensional
model using ground states gave the same results as the previously used
replica method. This makes us confident that the latter method also
yields sensible results in three dimensions where the ground state is
unknown. In particular it is worth noting that the two-replica
correlation function is sensible to the presence of a large number of
nearly degenerate states, whereas the ground state overlap measure
only correlations of the spin configurations with the global minimum
of the energy function.

Thus, having the concept of many pure states in spin glasses in mind
this, this observation seems to be pretty important: in two dimensions
the non-equilibrium dynamics is surprisingly similar to the
one-dimensional case, where no frustration is present, and if the same
observation could be made in the three-dimensional EA-model, it
behaves like a disguised ferromagnet with the ferromagnetic
ground state replaced by some other state and the magnetization as an
order parameter replaced by the global overlap with this state and
identical to the (now trivial) EA-order parameter. The dynamics is
slowed down by the presence of many metastable states and (free)
energy barriers between them that obey an extremely broad distribution
--- but otherwise nothing dramatical might be observable: neither
chaos in spin glasses \cite{fh,overlap} nor aging in various
asymptotic regimes \cite{fdt,skage}.

Of course the latter remarks are speculative. However, this is
essentially the picture that emerges from our simulations and the
results that we have reported in this paper. One conclusion might be
that the characteristics of the spin glass dynamics, which
discriminates itself from a slow quasi-ferromagnetic domain growth,
become observable only on much larger length (and time) scales than
those attainable by Monte Carlo simulations. The same, as we think, is
also true with respect to experiments: although the number of decades
(in time) that can be explored in a single experiments is
approximately 5 (e.g.\ from 1 to $10^4$ seconds in a typical aging
setup \cite{refregier,vincent,granberg}) and thus even smaller than in
our simulations, the time scales themselves can vary over a much
broader range. However, assuming a microscopic time of $10^{-12}$
seconds, which we deliberately identify for the moment with 1 Monte
Carlo step in our simulations, the above mentioned experimental time
scale could only be reached by performing $10^{12}$ to $10^{16}$
MC-steps, which looks hopeless at the moment. The length scales that
are physically relevant are indeed comparable, though, simply for the
reason of the extremely sluggish domain growth: if the latter would be
logarithmic, as proposed by the droplet theory \cite{fh}, the
correlation length reached in the experiments will be not much larger
(depending on the exponent $\psi$) than twice the correlation length
of the Monte Carlo simulations. We do not think that the physics is
much different then.

The lesson that we have to learn from it is the following: Most of the
existing theories for the non-equilibrium dynamics of spin glasses,
which are claimed to be valid on {\it asymptotic} time scales, seem to
be inappropriate for the description of the physics on {\it
intermediate} length scales that are relevant for the experiments and
for the results obtained in this paper. We have suggested a modified
scaling picture, whose basic assumption is a logarithmic growth of
energy barriers as a function of the domain size, which describes
consistently the physics of the two- and three-dimensional spin glass
models on the length scales we were able to explore. Obviously a more
complete theory is asked for, which might reveal the deeper reason for
these ad hoc scaling assumption, which we leave open as a challenge to
future research.

\section[*]{Acknowledgement}

We would like to thank the Center of Parallel Computing (ZPR) in K\"oln
and the HLRZ at the Forschungszentrum J\"ulich for the generous allocation of
computing time on the transputer cluster
Parsytec--GCel1024 and on the Intel Paragon XP/S 10. This work was performed
within the SFB 341 K\"oln--Aachen--J\"ulich.

\begin{table}
\vskip0.3cm

\hskip-0.4cm
\bt{|ll|c|cc|}
\hline
\rule[-3mm]{0cm}{0.9cm}
 & & Droplet-model \cite{fh} & MC-sim.\ & \\ \hline
\rule[-0mm]{0cm}{0.6cm}
Energy barrier & {\bf B} & $\Delta L^\psi$ & $\Delta(T)\log L$ & $\bullet$ \\
\rule[-0mm]{0cm}{0.6cm}
Activated dynamics & {\bf $\tau$}
                   & $\tau\sim\exp\,B/T$ & $\tau\sim\exp\,B/T$ &  \\
\rule[-0mm]{0cm}{0.6cm}
Domain size & {\bf L(t)} & $\left(\frac{T}{\Delta}\log\,t\right)^{1/\psi}$
                       & $t^{\alpha(T)}$ & ($\bullet$)  \\
\rule[-0mm]{0cm}{0.6cm}
Remanent Magnetization & {\bf M$_{TRM}$}
                     & $\left(\frac{T}{\Delta}\log\,t\right)^{-\delta/\psi} $
                       & $t^{-\lambda(T)}$ & $\bullet$  \\
\rule[-0mm]{0cm}{0.6cm}
Aging & {\bf C(t,t$_w$)}
      & $\overline{C}\left(\frac{\log(t/\tau)}{\log(t_w/\tau)}\right)$
      & $\tilde{C}\left(\frac{t}{t_w}\right)$ & $\bullet$  \\
\rule[-0mm]{0cm}{0.6cm}
Asymptotic decay & {\bf t$\gg$t$_w$} & $(\log\,t)^{-\delta/\psi}$
                                     & $t^{-\lambda(T)}$ & $\bullet$  \\
\rule[-3mm]{0cm}{0.9cm}
                 & {\bf t$\ll$t$_w$} & $(\log\,t)^{-\theta/\psi}$
                 & $t^{-x(T)}$  & ($\bullet$)  \\
\hline
\et
\vskip0.5cm
\caption{Predictions of different scaling assumptions.
The bullets indicate that the Monte Carlo simulations
confirm the corresponding prediction in the last colomn.  A
bullet with brackets means that the numerical data can also be
interpreted according to the predictions of the droplet model.
Concluding the hypothesis (\protect{\ref{logbarrier}}) seems to give a more
consistent
description of experimental and numerical results on aging in spin
glasses presented so far.}
\label{tab1}
\end{table}

\newpage

\newpage

\newpage

\newpage

\newpage

\newpage

\newpage

\newpage

\newpage

\newpage

\newpage

\newpage

\end{document}